\documentclass[pre,twocolumn]{revtex4} 
\usepackage{amsmath,epsfig}

\newcommand{\be}{\begin{equation}}
\newcommand{\ee}{\end{equation}}
\newcommand{\bp}{\begin{picture}}
\newcommand{\ep}{\end{picture}}
\newcommand{\ba}[1]{\begin{array}{#1}}
\newcommand{\ea}{\end{array}}
\newcommand{\bea}{\begin{eqnarray}}
\newcommand{\eea}{\end{eqnarray}}

\begin{document}

\title{
Capillary Condensation and Interface Structure
of a Model Colloid-Polymer Mixture in a Porous Medium
}
\author{Paul P.\ F.\ Wessels}
\email{wessels@thphy.uni-duesseldorf.de}
\affiliation{Institut f\"{u}r Theoretische Physik II,
Heinrich-Heine-Universit\"{a}t 
D\"{u}sseldorf,
Universit\"{a}tsstra{\ss}e 1,
40225 D\"{u}sseldorf, Germany}

\author{Matthias Schmidt}
\altaffiliation[On leave from:]{
Institut f\"{u}r Theoretische Physik II,
Heinrich-Heine-Universit\"{a}t 
D\"{u}sseldorf,
Universit\"{a}tsstra{\ss}e 1,
40225 D\"{u}sseldorf, Germany}
\affiliation{Debye Institute, Utrecht University, 
Princetonplein 5, 3584 CC Utrecht, The Netherlands}

\author{Hartmut L\"{o}wen}
\affiliation{Institut f\"{u}r Theoretische Physik II,
Heinrich-Heine-Universit\"{a}t
D\"usseldorf,
Universit\"{a}tsstra{\ss}e 1,
40225 D\"{u}sseldorf, Germany}

\date{\today}

\begin{abstract}
We consider the Asakura-Oosawa model of hard sphere colloids and ideal
polymers in contact with a porous matrix modeled by immobilized
configurations of hard spheres.  
For this ternary mixture a
fundamental measure density functional theory is employed, where the
matrix particles are quenched and the colloids and polymers are
annealed, i.e.\ allowed to equilibrate.  
We study capillary condensation of the mixture in a tiny sample of
matrix as well as demixing and the fluid-fluid interface
inside a bulk matrix.
Density profiles normal to the interface and surface tensions
are calculated and compared to the case without matrix.
Two kinds of matrices are
considered: (i) colloid-sized matrix particles at low packing
fractions and (ii) large matrix particles at high packing
fractions. 
These two cases show fundamentally different behavior and should
both be experimentally realizable.
Furthermore, we argue that capillary condensation of a colloidal suspension
could be experimentally accessible.
We find that in case (ii), even at high packing fractions, 
the main effect of the matrix is to exclude volume and, to high accuracy, 
the results can be mapped onto those of the same system without matrix via a simple
rescaling.
\end{abstract}



\maketitle

\section{Introduction}
\label{sec:intro}

Bringing a fluid in contact with a porous medium has a profound
influence on its characteristics and phase behavior
\cite{evans90,gelb99}.  Due to abundance of surfaces and their
necessary proximity, surface-fluid interactions as well as capillarity
effects play a prominent role.  
Moreover, the system may be trapped in locally stable states, and its behavior
governed by hysteresis.
Apart from the above fundamental questions, the study of adsorbates in porous
media is also of great interest in applied fields
ranging from industrial and geophysical to biomedical 
and pharmaceutical systems~\cite{gelb99,smith98}.

Many natural porous materials are tremendously complex on a
microscopic scale: irregularly shaped pores build a connected void
space that percolates throughout the sample~\cite{hilfer00,mecke02a}. 
In contrast, to
facilitate systematic studies, one often relies on model pores like
slit-like, cylindrical or spherical pores (see~\cite{evans90,gelb99}
and Refs.\ therein).   
The pore is then described
conveniently in terms of a single parameter, its size.  A different
class of idealized system makes use of immobilized arrangements of
fluid particles (i.e.\ a quenched hard sphere fluid) to model a porous
medium 
(see~\cite{gelb99} and Refs.\ therein). 
In turn, this is characterized through its density and the size
of the spheres.  However, the
relevant difference to idealized pores is the presence of {\em random}
confinement.

The study of porous media has been focused so far mainly on atomic liquids.  
In a colloidal fluid, length and time scales are much larger,
facilitating e.g.\ studies in real space and time~\cite{lowen01}.  
We believe that the use of
colloidal suspensions as model systems to study the behavior of
adsorbates in porous media can be as beneficial as their use to study
many other phenomena in condensed matter.
However, the experimental challenge lies in constructing three-dimensional
porous media suitable for colloidal suspensions.

Colloidal porous media in 2D have been prepared by Cruz de Le\'{o}n et.\ al.~\cite{cruzdeleon98,cruzdeleon99}
by confining a suspension of large colloids between parallel glass plates.
Then, these served as a porous matrix to a fluid of smaller particles
which remained mobile and of which they measured the structure and effective potentials.
To our knowledge, no experiment similar in spirit has been performed in three dimensions to
date.
On the other hand, Kluijtmans et.\ al.\ constructed 3D porous glasses of silica spheres~\cite{kluijtmans97,kluijtmans99} 
and silica rods~\cite{kluijtmans00}, but studied the dynamics of isolated tracer colloids in these media.
Wero\'{n}ski et al.\ studied transport properties in porous media of glass beads~\cite{weronski03}.
Still, such glassy  arrangements of spherical colloids are a direct candidate for porous media 
suitable for colloidal suspensions.
Sediments of large and heavy colloids as used in Refs~\cite{kluijtmans97,kluijtmans99,weronski03}
could be brought in contact with a suspension of smaller density-matched (to the solvent) colloids
of which the local structure could be determined~\cite{cruzdeleon98,cruzdeleon99}.
However, the size ratio of the two species is a crucial control parameter: 
It has to be large enough ($\gtrsim 10$) such that the small particles can penetrate 
the void space, but should still be small enough such that no complete separation of length scales occurs. 
Another way to realize such porous media would be to use laser tweezers.
In a binary colloid mixture of which one of the species possesses the same index of refraction 
as the solvent (via index-matching) and the other type has a higher index of refraction
the second species could be trapped while the first would still remain mobile.
Using multiple traps at random positions in space (mimicking a fluid) one could then realize a
model porous matrix~\cite{hoogenboom02}. 
The advantages of this method are the accessibility of very low matrix packing fractions and
the full control of the confinement.  
However, the number of trapped colloids in such setups is typically limited to the order of 100 --
probably too little to approach real macroscopic porous media, but in the
right regime to be able to compare to computer simulations, where similar numbers are accessible.
The crucial advantage of these setups over the use of ``natural'' porous media is their model 
character arising from the use of well-defined monodisperse matrix spheres,
while these still possess the essential features of random confinement and
a highly interconnected void structure. 

One prominent phenomenon that is induced by confinement is capillary
condensation: A liquid inside the porous medium is in equilibrium
with its vapor outside the medium.  In order for a substance to
phase separate into a dense liquid and a dilute gas phase a
sufficiently long-ranged and sufficiently strong attraction between
the constituting particles is necessary.  It is well-known that the
addition of non-adsorbing polymers to a colloidal dispersions induces
an effective attraction between the colloids.  The polymer coils are
depleted from a shell around each colloid and overlap of these
(depletion) shells generates more available volume to the polymers
yielding an effective attraction between the colloids.  Consequently,
these colloid-polymer mixtures may separate into a colloid-poor (gas)
phase and a colloid-rich (liquid) fluid~\cite{poon02}.

The most simplistic theoretical model that has been applied for the
study of such colloid-polymer mixture is the Asakura-Oosawa (AO) 
model~\cite{asakura54,asakura58,vrij76} that
takes the colloids to be hard spheres and the polymers to be ideal
spheres that are excluded from the colloids.  
The bulk phase behavior of this model was studied with a variety of techniques, 
like effective potentials~\cite{gast83, dijkstra99}, free volume theory~\cite{lekkerkerker92}, 
density functional theory (DFT)~\cite{schmidt00cip,schmidt02cip} and 
simulations~\cite{dijkstra99,bolhuis02phasediag,dijkstra02swet}. 
Recent work has also been devoted to inhomogeneous situations, i.e.\ the free interface
between demixed fluid phases~\cite{vrij97,brader00,brader02swet,brader01phd},
the adsorption behavior at a hard wall, where in particular a novel type of entropic
wetting was found~\cite{brader02swet,brader01phd,dijkstra02swet}
and the behavior in spatially periodic external potentials~\cite{goetze03}.
The surface tension between demixed colloid-polymer systems has been measured experimentally
and established to be much lower than for atomic systems~\cite{dehoog99,dehoog99II,dehoog01phd,aarts03swet}.
Further, recent experiments confirm wetting of the colloid-rich liquid at a hard
wall~\cite{dirkpriv03,wijting03}.

DFT~\cite{evans92} can be used in two ways to treat adsorbates in
porous media. The first is the (conceptually) straightforward approach
via treating the porous medium as an external potential (see e.g.\
Refs.~\cite{frink02,frink03one,frink03two}) and to solve for the one-body
density distributions of the fluid species. 
Those can be complicated spatial distributions, hence this approach is computationally
demanding, 
but also yields information on out-of-equilibrium behavior
like hysteresis in ad- and desorption curves
\cite{kierlik01,kierlik02,rosinberg02}.

A recently proposed alternative is to describe the quenched
component on the level of its one-body density distribution~\cite{schmidt02pordf}. 
Following the fundamental measure theory (FMT) 
of hard spheres~\cite{rosenfeld89,RSLTlong,tarazona00}, an explicit scheme was obtained 
to generate an approximate excess free energy for (not necessarily additive) hard-sphere mixtures in contact
with hard-sphere matrices~\cite{schmidt02pordf}.
Applied to the AO model, the results were compared with those from solving the so-called
replica Ornstein-Zernike (ROZ) equations~\cite{maddenglandt,madden92,givenstell,givenstell94} and found to be 
in good agreement~\cite{schmidt02aom}.
Meanwhile, this quenched-annealed (QA) DFT has been compared to computer simulations~\cite{schmidt03porz} 
and extended to hard-rod matrices~\cite{schmidt03porsn} and lattice fluids~\cite{lafuente02prl,schmidt03lfmf}.
FMT in combination with mean field theory has also been applied to fluids 
inside model pores~\cite{ravikovitch01,figueroa03}.

In this article, we revisit the AO model in contact with a hard-sphere matrix
using the QA DFT of Refs.~\cite{schmidt02pordf,schmidt02aom}.
We study capillary condensation in a tiny sample of matrix as well as the fluid-fluid
interface inside a bulk matrix.
For both these phenomena, we distinguish two cases of matrices: (i) matrix particles having the same
size as the colloids and (ii) where they are much larger.
These correspond to the two possible experimental realizations we discussed earlier in the introduction,
but also serve as representative cases because their behavior is fundamentally different.
Concerning capillary condensation, we focus on the possible experimental realization and
consider a bulk mixture in contact with in tiny sample of matrix.
Furthermore, we elaborate if and how capillary condensation could be observable in such experiments.
Concerning the fluid-fluid interface, we study the interfacial profiles
as well as the surface tensions inside the matrix.
For the case of small matrix particles (i), we determine the nature of decay (monotonic or periodic) 
of the interfaces which we compare with the bulk pair correlations.
For the case of large matrix particles (ii), we observe a simple rescaling of the bulk
as well as the interface results with respect to the case without matrix.
Inhomogeneous situations like the fluid-fluid interface, are treated within QA DFT
in a direct fashion, in contrast to e.g.\ the ROZ equations.
Fluid-fluid interfaces have been studied before in Lennard-Jones systems in contact with porous
media using the Born-Green-Yvon equation as well as computer simulations~\cite{trokhymchuk98,reszko-zygmunt02}
and we briefly compare to results of our profiles.

The paper is organized as follows. 
In Sec.~II we define our theoretical model explicitly. 
The QA DFT approach is reviewed in Sec.~III, and the results are presented in Sec.~IV.
We first consider capillary condensation in a tiny sample
and then demixing, the interfacial profiles and tensions inside a matrix.
We conclude with a discussion in Sec.~V.

\begin{figure}[t]
\centering
\epsfig{figure=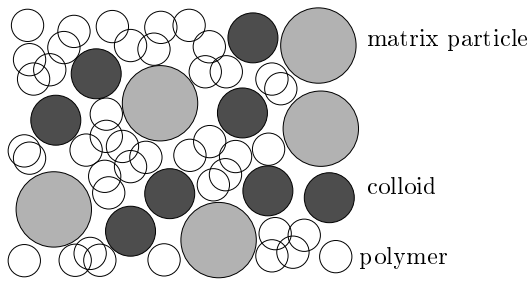,width=7.cm}
\caption{\small Sketch of the ternary mixture of mobile colloids (dark), mobile polymers 
(transparent) and immobile matrix particles (grey).
The polymer coils can freely overlap.
There are three model parameters, i.e.\ the packing fraction of matrix particles, $\eta_m$
and two size ratios, $q=R_p/R_c$ and $s=R_m/R_c$ where $R_i$ are is the radius of particles of species $i$.
The packing fractions of colloids and polymers, $\eta_c$ and $\eta_p$ respectively,
are the thermodynamic parameters.
}
\label{fig:ternarymixture}
\end{figure}

\section{Model}

We consider a three-component mixture of colloids (denoted by \emph{c}), polymers (\emph{p})
and immobile matrix particles (\emph{m}).
Each of these particles are spherical objects with radii $R_i$ and $i=c,p,m$
and corresponding number densities $\rho_i=N_i/V$ where $N_i$ is the total number
of molecules of species $i$ and $V$ is the system volume.
All of these components are modeled as hard bodies meaning they cannot overlap
but otherwise do not interact with each other, except for the polymer-polymer interaction,
which is taken to be ideal, see Fig.~\ref{fig:ternarymixture}.
Consequently when $r$ is the mutual distance, the pair potentials
become
\begin{multline}
u_{ij}(r)=\left\{ 
\ba{cc} 
\infty &\mbox{if $r<R_i+R_j$}
\\ 
0&\mbox{if $r\geq R_i+R_j$}
\ea
\right.
\\ 
\mbox{for $i,j=c,p,m$ except for $i=j=p$}
\end{multline}
and concerning the polymer-polymer interaction, this simply becomes
\be\ba{cc}
u_{pp}(r)=0&\mbox{for all $r$}
\ea .
\label{eq:idealpotential}
\ee
As all interactions are either hard-core or ideal, the (phase) behaviour is governed by
entropic (packing) effects and the temperature $T$ does not play a role.
The only thermodynamic parameters are the colloid and polymer packing fractions 
$\eta_c=4\pi R_c^3\rho_c/3$ and $\eta_p=4\pi R_p^3\rho_p/3$, respectively.
The remaining model parameters are two size ratios $q=R_p/R_c$ and $s=R_m/R_c$
and the packing fraction of matrix particles $\eta_m=4\pi R_m^3\rho_m/3$. 
It has to be mentioned that due to the fact that the polymers can freely overlap, 
the ``polymer packing fraction'' can easily be larger than one (Fig.~\ref{fig:ternarymixture}).
The mixture of hard spheres with these last-mentioned ideal polymers (i.e.\ without the matrix
particles) is called the Asakura-Oosawa (AO) mixture~\cite{asakura54,asakura58}.

\section{Density Functional Theory}

\subsection{Zero-Dimensional Limit}

In this subsection we derive the zero-dimensional (0D) Helmholtz free energy for
the three-component system of the AO colloid-polymer mixture in contact with quenched hard spheres.
This 0D free energy is used as an input to construct the fundamental measure theory
in the next subsection
Here, we give only a brief derivation, a more extensive version with more comments
is given in Refs.~\cite{schmidt02pordf,schmidt02aom}.
The essential ingredient is that we need to perform the so-called
``double average'' which refers to the statistical average over all
fluid configurations and subsequently over all matrix realizations.
To that end we consider a 0D cavity which either does or does not contain
a matrix particle.
Hence, the 0D partition sum is that of a simple hard sphere fluid,
\be
\bar{\Xi}_m = 1+ \bar{z}_m,
\label{eq:matrix0dsum}
\ee
where $\bar{z}_m=\zeta \exp(\beta \bar{\mu}_m)$ is the fugacity of the hard spheres.
Further, $\beta=1/k_{\rm B} T$ with $k_{\rm B}$ Boltzmann's constant and $\bar{\mu}_m$ the 
chemical potential.
The irrelevant prefactor $\zeta$ scales with the vanishing volume of the cavity
but has no effect to the final free energy and will not be discussed further.
In general, we use an overbar to refer to quantities of 0D systems.
With the grand potential, $\beta\bar{\Omega}_m=-\ln\bar{\Xi}_m$, the average
number of matrix particles is $\bar{\eta}_m=-\bar{z}_m \partial 
\beta\bar{\Omega}_m/\partial \bar{z}_m=\bar{z}_m/(1+\bar{z}_m)$.

Next, we consider the colloid-polymer mixture in contact with the matrix in zero dimensions.
If the cavity is occupied by a matrix particle, no colloid or polymer can be present.
On the other hand, if there is no matrix particle, it can either be empty,
occupied by a single colloid or an arbitrary number of polymers.
Hence,
\be
\bar{\Xi} =\left\{ \ba{ll} 
1&\mbox{matrix particle in cavity}\\ 
\bar{z}_c+\exp(\bar{z}_p) & \mbox{no matrix particle in cavity}
\ea\right. ,
\label{eq:0dsum}
\ee
where $\bar{z}_c$ and $\bar{z}_p$ are the colloid and polymer fugacities respectively. 
Then, the contribution $-\ln\bar{\Xi}$ to the grand potential should contain the 
appropriate statistical weight for each of the cases, i.e.\ $\bar{z}_m/\bar{\Xi}_m$ 
for the first and  $1/\bar{\Xi}_m$ for the second,
\be
\beta\bar{\Omega} =
-\frac{\ln (\bar{z}_c+\exp(\bar{z}_p))}{1+\bar{z}_m}.
\label{eq:0dgrandpot}
\ee
Average particle numbers are again readily obtained via $\bar{\eta}_i=-\bar{z}_i \partial 
\beta\bar{\Omega}/\partial \bar{z}_i$ for $i=c,p$ (not for $m$).
The Helmholtz free energy can then be calculated using a standard Legendre transformation,
$\beta\bar{F}=\beta\bar{\Omega}+\sum_{i=c,p}\bar{\eta}_i\ln(\bar{z}_i)$
and we obtain for the excess part, 
$\beta \bar{F}_{\rm exc}=\beta \bar{F}-\sum_{i=c,p}\bar{\eta}_i[\ln(\bar{\eta}_i)-1]$,
\begin{multline}
\beta \bar{F}_{\rm exc}(\bar{\eta}_c,\bar{\eta}_p;\bar{\eta}_m)
= (1-\bar{\eta}_c-\bar{\eta}_p-\bar{\eta}_m)\ln (1-\bar{\eta}_c-\bar{\eta}_m)
\\ +\bar{\eta}_c -(1-\bar{\eta}_m)\ln (1-\bar{\eta}_m).
\label{eq:0dfreeenergy}
\end{multline}
This result can be shown to be equal from that which would be obtained using the
so-called ``replica trick''~\cite{givenstell94}.

\subsection{Fundamental Measure Theory}

Fundamental measure theory (FMT) is a nonlocal density functional theory, in which 
the excess part of the {\em three-dimensional} free energy 
$F_{\rm exc}$ is expressed as a spatial integral over the free energy density $\Phi$,
\begin{equation}
\beta F_{\rm exc}[\{\rho_{i}(\mathbf{r})\}]
=\int d\mathbf{r} \Phi(\{n^{i}_{\nu}(\mathbf{r})\}).\label{eq:freeenergydensity}
\end{equation}
This free energy density in turn is assumed to depend on the full set of 
weighted densities $\{n^{i}_{\nu}(\mathbf{r})\}$,
\begin{align}
n^{i}_{\nu}(\mathbf{r})&=\int d\mathbf{r}' w^{i}_{\nu}(\mathbf{r}-\mathbf{r}')
\rho_{i}(\mathbf{r}'),\label{eq:weighteddensity} \\
\nonumber
&=(w^{i}_{\nu}\otimes\rho_{i})(\mathbf{r})
\end{align}
which are convolutions (denoted with $\otimes$) with the single-particle distribution functions $\rho_{i}(\mathbf{r})$
for species $i=c,p,m$.
The weight functions are obtained from the low-density limit where the virial series
has to be recovered,
\begin{equation}
\begin{array}{ll}
w^i_3(\mathbf{r})=\theta(R_i-r) & \\
w^i_2(\mathbf{r})=\delta(R_i-r) 
& \mathbf{w}^i_{v2}(\mathbf{r})=\delta(R_i-r)\mathbf{r}/r \\
w^i_1(\mathbf{r})=\delta(R_i-r)/(4\pi r) &
\mathbf{w}^i_{v1}(\mathbf{r})=\delta(R_i-r)\mathbf{r}/(4\pi r^2)\\
w^i_0(\mathbf{r})=\delta(R_i-r)/(4\pi r^2) &
\end{array} \label{eq:weights}\end{equation}
with again $i=c,p,m$ being one of the three components $\theta$ the Heaviside function and 
$\delta$ the Dirac delta function.
There are four scalar weight functions, with 3 to 0, corresponding respectively to 
the volume of the particles, 
the surface area, the mean curvature and the Euler characteristic and these are
the so-called ``fundamental measures'' of the sphere.
The two weights on the right-hand side of Eq.~\ref{eq:weights} are vector quantities.
Often a seventh tensorial weight is used in the context of freezing but this will not be used 
here~\cite{tarazona00,schmidt00cip,schmidt02cip}.
The dimensions of the weight functions $w^i_{\nu}$ are $({\rm length})^{\nu-3}$.

Then, the sole approximation made is that $\Phi$ is taken to be a {\em function}
of the weighted densities $n^i_{\nu}(\mathbf{r})$ whereas most generally one would
expect this to be a {\em functional} dependence.
This approximation totally sets the form of $\Phi$ and following Refs.~\cite{schmidt02pordf,schmidt02aom,schmidt02cip}
we give the expression for $\Phi=\Phi_1+\Phi_2+\Phi_3$ in terms of the zero-dimensional 
free energy derived in the previous subsection,
\begin{align}
\Phi_1&=\sum_{i=c,p,m} n_0^i \varphi_i(\{n_3^l\}),\label{eq:phi1}\\
\Phi_2&=\sum_{i,j=c,p,m}\left( n_1^i n_2^j 
-\mathbf{n}_{v1}^i\cdot\mathbf{n}_{v2}^j \right) \varphi_{ij}(\{n_3^l\}),\label{eq:phi2}\\
\Phi_3&=\frac{1}{8\pi}
\sum_{i,j,k=c,p,m}\left( {\textstyle \frac{1}{3}} n_2^i n_2^j n_2^k 
- n_2^i \mathbf{n}_{v2}^j\cdot\mathbf{n}_{v2}^k
\right) \varphi_{ijk}(\{n_3^l\})
\label{eq:phi3}
\end{align}
with
\be
\varphi_{i_1 \ldots i_t}(\{\bar{\eta}_j\})=
\partial^t \beta \bar{F}_{\rm exc}(\{\bar{\eta}_j\}) 
/\partial \bar{\eta}_{i_1}\ldots \partial \bar{\eta}_{i_t}.
\label{eq:varphi}
\ee
All $\varphi_{i_1 \ldots i_t}$ of which more than one indices equal $\emph{p}$ 
are zero due to the form of $\beta \bar{F}_{\rm exc}$.
Together, Eqs.~(\ref{eq:0dfreeenergy}) to~(\ref{eq:varphi}) constitute the
excess free energy functional for this QA system.

\subsection{Minimization}

Having constructed the excess free energy, we can now immediately move on to the
the grand-canonical free energy functional of the colloid-polymer mixture in contact with a matrix, 
\begin{multline}
\Omega[\rho_{c}(\mathbf{r}),\rho_{p}(\mathbf{r});\rho_{m}(\mathbf{r})]= 
F_{\rm exc}[\rho_{c}(\mathbf{r}),\rho_{p}(\mathbf{r});\rho_{m}(\mathbf{r})]+\\ +
k_{\rm B} T \sum_{i= c,p}\int d\mathbf{r} \rho_i (\mathbf{r}) 
\left[\ln\left(\rho_i (\mathbf{r})\Delta_i\right)-1\right] + \\ +
\sum_{i= c,p}\int d\mathbf{r} \rho_i (\mathbf{r}) \left[ V_i(\mathbf{r})-\mu_i\right]
\label{eq:grandfunctional}
\end{multline}
Here, $\Delta_i$ is the ``thermal volume'' which is the product of the relevant de Broglie
wavelengths of the particles of species $i$.
Further, $\mu_i$ is the chemical potential and $V_i$ is the (external) potential acting
on component $i$.
In this paper, we study bulk phase behaviour and the free fluid-fluid interfaces so we use 
$V_i=0$.
The equilibrium profiles are the ones that minimize the functional,
\be\ba{ccc}{\displaystyle
\frac{\delta \Omega}{\delta \rho_c(\mathbf{r})}=0}
&\mbox{and}&{\displaystyle\frac{\delta \Omega}{\delta \rho_p(\mathbf{r})}=0}
\ea .\label{eq:minimization}\ee
This yields the Euler-Lagrange or stationarity equations ($i=c,p$),
\be
\rho_{i}(\mathbf{r})=z_i \exp\left[
c_i^{(1)}\left(\{\rho_j (\mathbf{r})\} \right)
\right],\label{eq:eulerlagrange}
\ee
with $z_i=\Delta_i^{-1}\exp[\beta \mu_i]$ the fugacity of component $i$ and 
the one-particle direct correlation functions given by
\be
c_i^{(1)}(\mathbf{r})=-\beta \frac{\delta F_{\rm exc}[\{\rho_j 
(\mathbf{r})\}]}{\rho_i (\mathbf{r})} =
-\sum_{\nu}  \left(\frac{\partial \Phi} {\partial n^i_{\nu}}\otimes w^i_{\nu}\right)(\mathbf{r})
\label{eq:onedirectcorr}
\ee
Obviously, the functional is not minimized with respect to the matrix distribution 
$\rho_m(\mathbf{r})$ as this serves as an {\em input} profile.
In principle, as we are dealing with a quenched-annealed system in which the matrix is initially
(before quenching) a hard sphere fluid, $\rho_m(\mathbf{r})$ should still minimize the hard sphere 
functional~\cite{schmidt02pordf,schmidt02aom}.
However, as density functional theory allows us to generate any distribution $\rho_m(\mathbf{r})$ 
by applying any suited external potential (which we can then remove after quenching), we do not
need to go into the scheme of generating matrix profiles.
Moreover, in the present paper we use fluid distributions of the matrix particles which minimize
(at least locally) the hard sphere functional without external potential for any packing fraction.

\section{Results}

In this section, we show results of the effect of the hard sphere matrix on the AO
colloid-polymer mixture concerning capillary condensation in a tiny sample of matrix, and
phase behaviour and free fluid-fluid interfaces inside a bulk matrix.
Throughout this section, we distinguish between the colloid-sized matrix particles ($s=1$)
and the large matrix particles (we use $s=50$).
In the first case, as we will see, one is limited to small matrix packing fractions (up to $\eta_m$ of the
order of $0.2$) as for high packing
fractions the pores become too small for the colloids and the polymers to constitute a real fluid 
in the matrix.
For the large matrix particles, higher matrix packing fractions are accessible (up to $\eta_m=0.5$).
Finally, in all cases we use $q=0.6$, for which size ratio the AO model has a stable 
fluid-fluid demixing area (with respect to freezing, which we do not consider).

\subsection{Bulk Fluid Free Energy}

In the fluid phase, the densities are spatially homogeneous, 
and the constant distributions $\rho_i(\mathbf{r})=\rho_i$ solve the 
stationarity Eqs.~(\ref{eq:eulerlagrange}) and~(\ref{eq:onedirectcorr}). 
Therefore, we only have to integrate the weights over
space, $\int d\mathbf{r}w_{\nu}^i$, and the weighted densities become
\be
\begin{split}
n_3^i&=\eta_i,\\
n_2^i&=3\eta_i/R_i,\\
n_1^i&=3\eta_i/(4\pi R_i^2),\\
n_0^i&=3\eta_i/(4\pi R_i^3),\\
\mathbf{n}_{v2}^i&=\mathbf{n}_{v1}^i=0,\\
\end{split}
\label{eq:homoweighteddensities}
\ee
with $i=c,p,m$.
Substituting these expressions in the free energy density, Eq.~(\ref{eq:phi1}) to~(\ref{eq:phi3})
we obtain an analytical expression for the bulk excess free energy.
Defining the dimensionless bulk free energy density, $f=\beta FV_c/V$ with $V_c=4\pi R_c^3/3$ the
volume of a colloid, this becomes
\begin{multline}
f(\eta_c,\eta_p;\eta_m)=\eta_c\left(\ln\eta_c-1\right)+\frac{\eta_p}{q^3}\left(\ln\eta_p-1\right)
\\ +f_0(\eta_c,\eta_m)-\frac{\eta_p}{q^3}\ln\alpha(\eta_c,\eta_m)
\label{eq:bulkfreeenergydensity}
\end{multline}
with the last two terms being the excess free energy.
We have separated the excess free energy in two terms where $f_0$ the excess free energy density
of a fluid of hard spheres in contact with a hard sphere matrix and $\alpha$ is the
fraction of free volume for the polymers in the presence the hard sphere colloidal
fluid {\em and} the hard sphere matrix~\cite{lekkerkerker92,schmidt02cip}.
The expressions for $f_0$ and $\alpha$ are quite extensive and given in the appendix.
In going from Eq.~(\ref{eq:freeenergydensity}) to~(\ref{eq:bulkfreeenergydensity})
we have discarded two terms, $\eta_c \ln(\Delta_c/V_c)$ and $(\eta_p/q^3) \ln(\Delta_p/V_p)$,
linear in the colloid and polymer packing fractions.
These have no effect on the phase behaviour.
Due to the ideal interactions of the polymers, the excess free energy density is only linear in $\eta_p$
and the polymer fugacity becomes simply
\be
z_p V_p=\eta_p/\alpha (\eta_c,\eta_m).
\label{eq:polymerfugacity}
\ee
This relation is trivially invertible, so switching from system
representation (using $f(\eta_c,\eta_p;\eta_m)$) to the
polymer reservoir representation (in terms of $\tilde{\omega}(\eta_c,z_p;\eta_m)=f-\mu_p\eta_p/q^3$) 
is straightforwardly done.
Moreover, for zero packing fractions of colloids and matrix particles, the polymer
free volume fraction is trivial, $\alpha(0,0)=1$.
Consequently, the fugacity equals the packing fraction of polymers in the polymer 
reservoir, $z_pV_p=\eta_{p,r}$ (where there are no colloids and matrix particles), and
often, we use $\eta_{p,r}$ when referring to the fugacity.
Finally, we mention that in the absence of matrix particles, $\eta_m=0$, this theory is
equivalent to the free-volume theory for the AO model~\cite{lekkerkerker92,schmidt00cip,schmidt02cip}.

Concerning the fluid-fluid demixing, the spinodals are calculated in the canonical representation, by solving 
$\det[\partial^2 f/\partial\eta_i\partial \eta_j]=0$ with $i,j=c,p$, which can be done
analytically.
Binodals are determined by constructing the common tangents of the semi-grand potential
$\tilde{\omega} (\eta_c,\eta_{p,r};\eta_m)$ at fixed fugacity $\eta_{p,r}$.

When the matrix particles are very large, it is expected that the excluded volume effects
dominate over other (surface or capillary) effects.
In particular, if one considers only one infinitely large particle, still corresponding 
to a nonzero matrix packing fraction, one would expect normal bulk behavior of the mixture
as most of the mixture is ``far'' away from the matrix particle.
Equivalently, for very large matrix particles, the total volume
of the surrounding depletion layers, which are responsible for the surface effects,
compared to the actual volume occupied by the matrix particles
scales with $(4\pi(R_m+R_c)^3\rho_m/3-\eta_m)/\eta_m\propto 3/s$
for the colloids and $(4\pi(R_m+R_p)^3\rho_m/3-\eta_m)/\eta_m\propto 3q/s$ for the polymers,
and these both go to zero for $s\rightarrow\infty$.
However, in this limit, we still need to correct for the volume as this is partly
occupied by infinitely large matrix particles, i.e.\ $V\rightarrow (1-\eta_m)V$.
Indeed, applying $s\rightarrow\infty$ to the bulk free energy of 
Eq.~(\ref{eq:bulkfreeenergydensity}),
we re-obtain the bulk behaviour of the plain AO colloid-polymer mixture {\em without} matrix,
i.e.\ it can be shown that
\be
\lim_{s\rightarrow\infty} f(\eta_c,\eta_p;\eta_m)=
(1-\eta_m)f\left(\frac{\eta_c}{1-\eta_m},\frac{\eta_p}{1-\eta_m};0\right),
\label{eq:rescaling}
\ee
where the free energy density has to be rescaled as well.
This term can be considered to be the zeroth in a $1/s$-expansion of the free energy
of which higher order terms should correspond to effects due to
surfaces, capillarity, curvature etc.
However, because of the formidable form of the free energy it is a daunting task
to connect every term to a certain phenomenon and we leave this to future investigation.
It is worth mentioning that a power series in $1/s$ is only a simple model dependence.
In general, there can be non-analyticities, e.g.\ arising from wetting phenomena around
curved surfaces~\cite{evans03} of matrix particles.
\begin{figure}[t]
\centering
\epsfig{figure=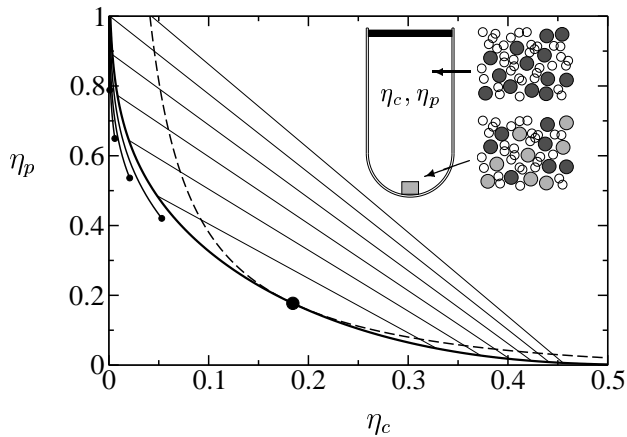, width=8.3cm}
\caption{\small 
Phase diagram of a bulk AO colloid-polymer mixture ($q=0.6$) 
in chemical contact with a tiny sample of porous material ($s=1$). 
Concerning the bulk mixture: the thick full curve is
the binodal, the dashed is the spinodal, the large filled circle (where they meet)
is the critical point and the straight (thin) lines are the tie lines connecting coexisting
state points.
The capillary lines (full curves) appear in the upper left (colloid-poor) part of the phase diagram 
and each terminates in a capillary critical point (small filled circles).
From lower-right to upper-left, the curves correspond to an increasing packing fraction of the matrix,
$\eta_m= 0.05, 0.1, 0.15, 0.2$ (the last one is practically on the vertical axis near $\eta_{p}=0.8$).
The inset shows a sketch of the setup: a test tube is sealed at the top and filled with
the colloid-polymer mixture (densities $\eta_c$ and $\eta_p$) and at the bottom lies
the tiny sample of porous matrix.
}
\label{fig:phasediagrams1tsr}
\end{figure}
\begin{figure}[t]
\centering
\epsfig{figure=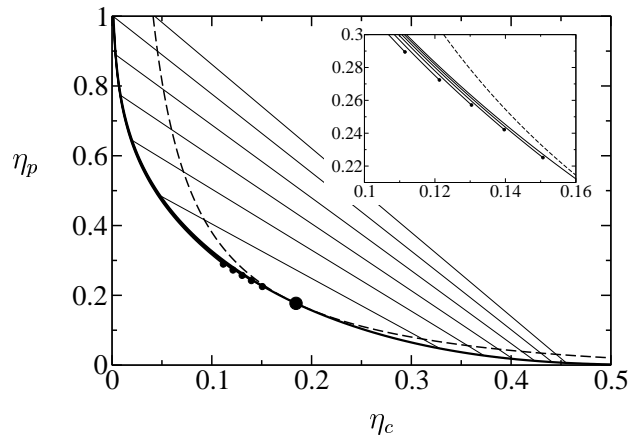, width=8.3cm}
\caption{\small 
Same as in Fig.~\ref{fig:phasediagrams1tsr}, but now for a sample of 
porous matrix with $s=50$.
Here the matrix packing fractions increase as follows: 
$\eta_m= 0.1, 0.2, 0.3, 0.4, 0.5$ (again from lower-right to upper-left).
The inset is the magnification of the area with the capillary critical points.
}
\label{fig:phasediagrams50tsr}
\end{figure}

\subsection{Capillary Condensation in a Tiny Sample of Porous Matrix}

A porous matrix of quenched hard spheres stabilizes the colloid-rich phase
with respect to the colloidal gas phase~\cite{schmidt02aom}.
This is called capillary condensation and it is due to the attractive depletion
potential between the colloids, which also acts between colloids and matrix particles.
In this subsection we present capillary condensation in a representation which is appropriate to compare with
experiments.
In an experimental setup one typically has a canonical ensemble, i.e.\ a test tube, of colloid-polymer
mixture.
By adding a tiny sample of porous material, the bulk mixture in the test tube acts as a
colloid-polymer reservoir to the sample, but vice versa, if the sample is small enough,
its state will not have any effect on that of the bulk mixture (see Fig.~\ref{fig:phasediagrams1tsr}(inset)).
In the colloid-poor (and polymer-rich) part of the phase diagram, on approaching
bulk coexistence, the conditions for coexistence in the porous sample are reached before
those in bulk, i.e.\ capillary condensation in the sample occurs.
Hence, capillary condensation appears as a line in the system representation terminating
in a capillary critical point.
This is shown in Figs.~\ref{fig:phasediagrams1tsr} and~\ref{fig:phasediagrams50tsr}
for the case of $s=1$ and $s=50$ respectively and various densities of the matrix.
The coexistence of the bulk colloid-polymer mixture appears in the usual system representation,
where tie lines connect coexisting states.
For each of the matrix densities, a capillary line runs along the bulk binodal in the colloid-poor
part of the phase diagram.

First, we determine the conditions for coexistence inside the matrix, i.e.\ we compute
the combinations of chemical potentials $\mu_{c,{\rm coex}}^{\rm porous}$ and $\mu_{p,{\rm coex}}^{\rm porous}$,
for which demixing occurs within the porous sample.
These are fixed by the chemical potentials of the bulk colloid-polymer mixture,
$\mu_c^{\rm bulk}$ and $\mu_p^{\rm bulk}$,
so solving
\begin{align}
\mu_c^{\rm bulk}(\eta_c,\eta_p)&= \mu_{c,{\rm coex}}^{\rm porous}, \label{eq:coextsr} \\
\mu_p^{\rm bulk}(\eta_c,\eta_p)&= \mu_{p,{\rm coex}}^{\rm porous} \nonumber
\end{align}
for $\eta_c$ and $\eta_p$, we obtain the capillary lines in the phase diagram in system representation.
The trend can be spotted from Figs.~\ref{fig:phasediagrams1tsr} and~\ref{fig:phasediagrams50tsr},
increasing the matrix packing fraction in the sample, the capillary line moves away from the bulk
binodal but at the same time the capillary critical point shifts away from the bulk critical point.
Qualitatively this applies to both the $s=1$- and $s=50$-cases.
However, in the $s=50$-case the capillary lines extend to much closer to the bulk critical point, but
they are hardly distinguishable from the bulk binodal.
Concerning the colloid-sized matrix particles, $s=1$, these capillary lines are well separated from the
bulk binodal, but the capillary critical points are located much deeper into the
colloidal gas regime.
\begin{figure}[t]
\centering
\epsfig{figure=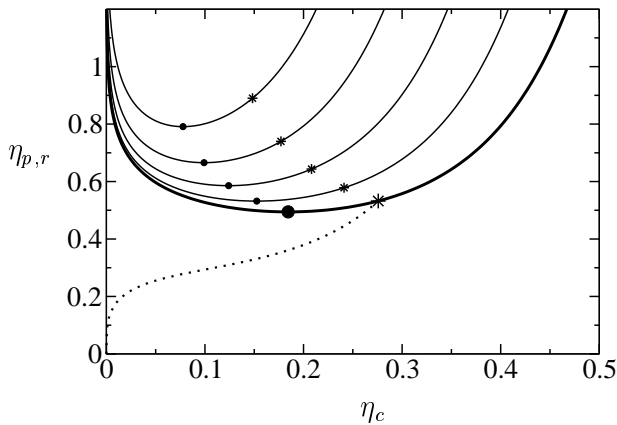,width=8.2cm}
\caption{\small Fluid-fluid binodals of an AO colloid-polymer mixture
($q=0.6$) inside a bulk porous matrix ($s=1$).
Tie lines connecting coexisting state point are not drawn but run horizontal.
The lower (thicker) curve is the result in the absence of any matrix, $\eta_m=0$.
For the other curves, the matrix packing fraction increases from bottom to top,
$\eta_m=0,0.05,0.1,0.15,0.2$.
The filled circles are the critical points (large, $\eta_m=0$).
The dotted line is the Fisher-Widom line for $\eta_m=0$ 
below of which the decay of correlations in the fluid is oscillatory
and above these are monotonic.
The point where the FW line hits the binodal is marked by a (large) star.
The FW lines for the other matrix packing fractions are not shown,
only their crossings with the binodals (small stars).
}
\label{fig:sgcphasediagrams1}
\end{figure}

Next, we briefly discuss the implications this has for possible experiments.
Focusing on the case of the large matrix particles, $s=50$, we take as an
example $\eta_m=0.5$.
In this case the difference in chemical potential at coexistence of
the mixture in bulk and inside the porous sample at constant polymer fugacity
is of the order, $\beta\Delta\mu_c^{\rm coex}\sim 0.1$, and it scales roughly with $1/s$.
This difference is very small and brings up the question if this (i.e.\ capillary condensation) 
is observable in experiments.
Typically, the effect of gravity is reduced
by density-matching the colloids with the solvent, i.e.\ canceling gravity by means of buoyancy.
However, this density-matching is never perfect, and the length scale $(\beta m_cg)^{-1}$ 
is a measure for its success (at infinity it is perfect).
Here $g$ is the gravitational acceleration and $m_c=(\rho_{\rm colloid}-\rho_{\rm solvent})V_c$ 
the effective mass of the colloid in solution, with $\rho_{\rm colloid}$ and 
$\rho_{\rm solvent}$ the mass densities inside the colloid and of the solvent respectively.
Therefore this length scale is strongly dependent on the colloid size, $(\beta m_cg)^{-1}\propto R_c^{-3}$,
and can range from micrometers (large colloids) to meters (small colloids) in experiments~\cite{dehoog01phd}.
Typically, polymers are much less sensitive to gravity as long as the solvent is good.
When there is coexistence inside a test tube, there is only real coexistence at the
liquid-gas interface whereas below and above the colloids have slightly different chemical potentials due to their
gravitational energy.
Consequently, moving upward from the interface, say $\Delta z$, the colloid chemical potential is 
$(\beta m_cg)\Delta z$ lower than at coexistence.
By placing the porous sample within $\Delta z^*=\beta\Delta\mu_c^{\rm coex}/(\beta m_cg)$ of the interface,
capillary condensation should take place.
Taking as an example, $\beta\Delta\mu_c^{\rm coex}\sim 0.1$ and $(\beta m_cg)^{-1}\sim 1$ meter, 
it becomes clear that, within the context of this (idealized) model, values of $\Delta z^*\sim 0.1$ meter should be accessible
in experiments, meaning that capillary condensation could in principle be observed.
Complete wetting of the large matrix spheres by the colloidal liquid
may preempt capillary condensation close to the bulk critical point.
Moving sufficiently far away from the critical point towards the dilute gas regime, 
the effects due to complete wetting should disappear while capillary condensation is retained.
\begin{figure}[t]
\centering
\epsfig{figure=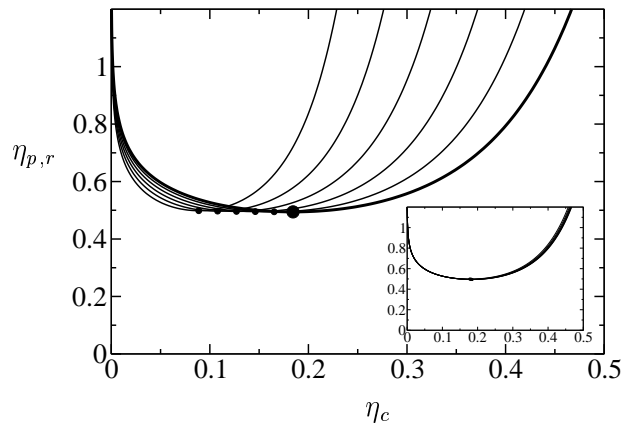,width=8.2cm}
\caption{\small Same as in Fig.~\ref{fig:sgcphasediagrams1}, but now
for $s=50$.
Matrix packing fractions increase from right to left, 
$\eta_m=0,0.1,0.2,0.3,0.4,0.5$.
Inset: same curves now rescaled, i.e.\ $\eta_{p,r}$ vs.\ $\eta_c/(1-\eta_m)$.
}
\label{fig:sgcphasediagrams50}
\end{figure}

\subsection{Phase Behaviour inside a Bulk Porous Matrix}

We now return to the full ternary mixture in bulk, i.e.\ where in the previous subsection, the matrix 
was only a tiny sample immersed in a large system of colloid-polymer mixture, in this and the 
following subsections we consider the colloid-polymer mixture in a system-wide matrix.
In this subsection, we revisit the demixing phase behavior which we need in the next subsections where we
study the fluid-fluid interface inside a matrix.
Fig.~\ref{fig:sgcphasediagrams1} is the phase diagram in the polymer-reservoir
representation for colloid-sized matrix particles, $s=1$, for various matrix densities.
Increasing the matrix packing fraction, there is less volume available
to the colloids and the critical point shifts to smaller colloid packing fractions.
At the same time, the porous matrix acts to keep the mixture ``mixed'' and therefore,
the critical point shifts to higher polymer fugacities.
For the case of  $s=1$, we can not go to much higher packing fractions than $\eta_m\sim 0.2$ as then
the critical fugacity shoots up dramatically to unphysically large values.
This may be partly due to the relatively large depletion shells around the matrix particles
which cause the pore sizes to become too small for the colloids and polymers to enter the matrix.
In case of large matrix particles ($s=50$, see Fig.~\ref{fig:sgcphasediagrams50}), the latter effect
is negligible and the pore sizes are always large enough.
Consequently, only the excluded volume remains and rescaling the binodals with $(1-\eta_m)$
is very effective practically mapping the binodals onto each other, Fig.~\ref{fig:sgcphasediagrams50}(inset).
This rescaling is unsuccessful for $s=1$ as can be directly seen from the fact that the
critical fugacities in Fig.~\ref{fig:sgcphasediagrams1} are different for each of the matrix densities.
\begin{figure}[t]
\centering
\epsfig{figure=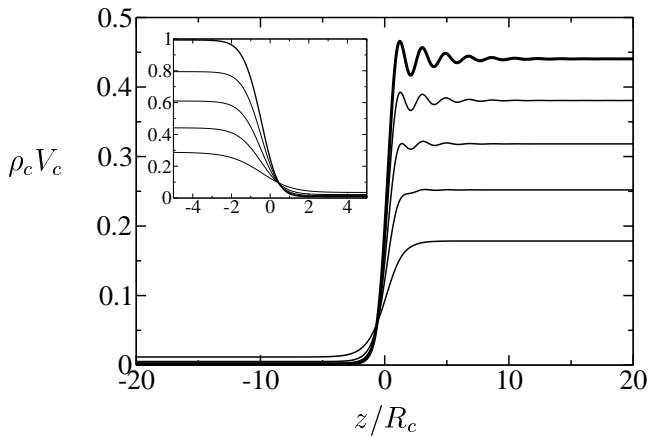,width=8.6cm}
\caption{\small Colloid density profiles ($V_c=\frac{4}{3}\pi R_c^3$) 
normal to the free fluid-fluid
interface for increasing matrix packing fractions at fixed polymer
fugacity.
Parameters are: $q=0.6$, $s=1$ and $\eta_{p,r}=1$ 
(see Fig.~\ref{fig:sgcphasediagrams1}).
The matrix packing fraction increases from top (thick profile, $\eta_m=0$) to bottom:
$\eta_m= 0, 0.05, 0.1, 0.15, 0.2$.
Inset: corresponding polymer profiles ($\rho_pV_p$ vs.\ $z/R_c$, with $V_p=\frac{4}{3}\pi R_p^3$) 
for the same values of the matrix 
packing fractions (also increasing from top to bottom).
}
\label{fig:colloidprofiless1}
\end{figure}

In addition, we have determined the nature of the asymptotic decay of pair correlations
of the fluid inside the matrix~\cite{evans94}.
These can either be monotonic or periodic and the corresponding regions in the phase diagram are 
separated by the Fisher-Widom (FW) line, at which both types of decay are equally long-range.
This line can be determined by studying the pole structure of the total correlation 
functions $h_{ij}$ in Fourier space~\cite{evans94}.
In the present case of QA systems, rather than using the usual Ornstein-Zernike equations,
one has to use the replica-Ornstein-Zernike (ROZ) equations~\cite{givenstell94}.
Neglecting correlations between the replicas, these are
\begin{align}
h_{mm}(\mathbf{r})&=c_{mm}(\mathbf{r})+\rho_m (c_{mm}\otimes h_{mm})(\mathbf{r}) \nonumber \\
h_{ij}(\mathbf{r})&=c_{ij}(\mathbf{r})+\sum_{t=c,p,m} \rho_t (c_{it}\otimes h_{tj})(\mathbf{r})
\end{align}
with $i,j=c,p,m$ except $i=j=m$.
Here, for $ij\neq mm$, the $c_{ij}(\mathbf{r})=-\delta^2 F_{\rm exc}/\delta \rho_i(\mathbf{r})\delta \rho_j(\mathbf{r})$
are the direct correlation functions for which we obtain analytic expressions by differentiating
Eq.~\ref{eq:freeenergydensity}.
The matrix structure is determined before the quench, so $c_{mm}$ and $h_{mm}$ are those of the
normal hard sphere fluid at density $\rho_m$ (Percus-Yevick-compressibility closure, see 
Refs.~\cite{schmidt02pordf,schmidt02aom}). 
This analysis follows closely that of Ref.~\cite{schmidt02cip} in which more details are given.
In view of our subsequent interface study, we focus on the point where the FW line meets the binodal.
In Fig.~\ref{fig:sgcphasediagrams1} ($s=1$), these are denoted by stars, and we observe that the shifts
due to the matrices follow the same trend as the critical points.
In case of $s=50$, we have not determined the FW lines, but there is no reason to expect
the simple rescaling of the case without matrix to fail in this case.
Furthermore, concerning the density profiles (in the next subsection, $s=50$), we stay well within the
oscillatory regime.
\begin{figure}[t]
\centering
\epsfig{figure=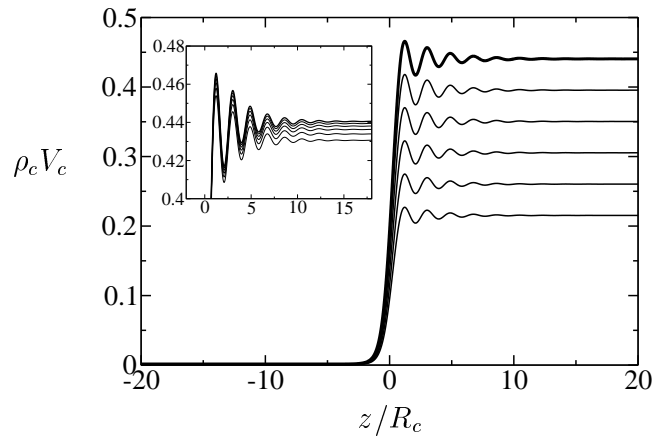,width=8.6cm}
\caption{\small 
Same as in Fig.~\ref{fig:colloidprofiless1} but now for $s=50$.
Other parameters are: $q=0.6$ and $\eta_{p,r}=1$ (see Fig.~(\ref{fig:sgcphasediagrams50})).
The matrix packing fractions increase from top (thick profile, $\eta_m=0$) to bottom:
$\eta_m= 0, 0.1, 0.2, 0.3, 0.4, 0.5$.
Inset: magnification of the {\em rescaled} profiles for the same curves, i.e.\
$\rho_cV_c/(1-\eta_m)$ vs.\ $z/R_c$
(where again, matrix packing fractions increase from top to bottom).
}
\label{fig:colloidprofiless50}
\end{figure}

\subsection{Fluid-Fluid Profiles inside a Bulk Porous Matrix}
\label{sec:profiles}

We have calculated density profiles at coexistence normal to the colloidal gas-liquid
interface.
In this case of planar interfaces, the density distribution is only a function of one
spatial coordinate $z$; i.e.\ $\rho_i(\mathbf{r})=\rho_i(z)$.
The only dependence on the other two degrees of freedom is in the weights
and this can be integrated out, to obtain projected weights,
$\tilde{w}^i_{\nu}(z)=\int dx dy w^i_{\nu}(\mathbf{r})$ (see e.g.~\cite{brader02rsa}).
The profiles are discretized and calculated via an iteration procedure, i.e.\ we insert
profiles on the right hand side of Eq.~(\ref{eq:eulerlagrange})
and then obtain new profiles on the left hand side, which are then reinserted on the right.
Using step functions as iteration seeds, this procedure converges in the (local) direction of 
the lowest free energy.
We normalize the densities as in bulk, i.e\ we plot $\rho_i(z)V_i$ so that 
$\rho_i(\pm\infty)V_i=\eta_i^{({\rm I,II})}$, with I and II referring
to the coexisting phases.
The zero of $z$ is set at the location of the interface, defined through
the Gibbs dividing surface of the colloids: $\int_{-\infty}^0 dz [\rho_c(z)-\rho_c(-\infty) ] 
+\int^{\infty}_0dz [\rho_c(z)-\rho_c(\infty) ]=0$.
\begin{figure}[t]
\centering
\epsfig{figure=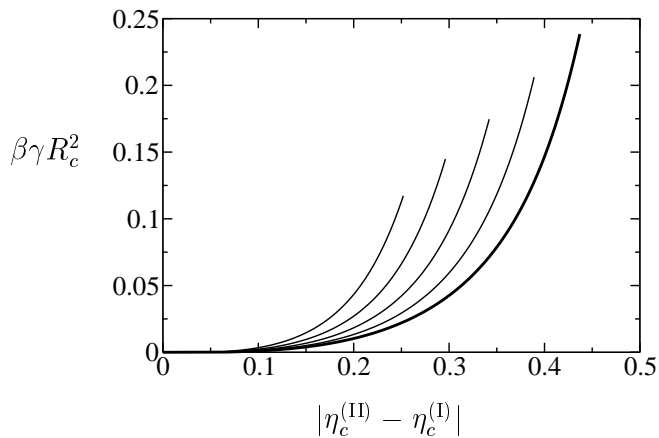,width=8.7cm}
\caption{\small Fluid-fluid surface tensions vs.\ the difference in colloid packing fractions
of the two fluid phases for $q=0.6$, $s=1$ and various values of the matrix packing fraction.
The matrix packing fractions increase from right ($\eta_m=0$, thick curve) to left,
$\eta_m= 0, 0.05, 0.1, 0.15, 0.2$.
Each curve is computed from the critical point, $\eta_{p,r}=\eta_{p,r}^{({\rm crit})}$
(where $\gamma =0$) until twice the critical fugacity, $\eta_{p,r}=2\eta_{p,r}^{({\rm crit})}$.
}
\label{fig:surfacetensions1}
\end{figure}

In Figs.~\ref{fig:colloidprofiless1} and~\ref{fig:colloidprofiless50}, we have plotted the
colloid profiles normal to the interface for $s=1$ and $s=50$, respectively.
Colloid profiles are shown for increasing densities of the matrix at fixed fugacity, $\eta_{p,r}=1$,
corresponding to the bulk binodals in Figs.~\ref{fig:sgcphasediagrams1} and~\ref{fig:sgcphasediagrams50}.
For the case of $s=1$, this means that, as the critical point shifts to higher fugacities, 
the profiles are effectively taken at fugacities closer to the critical value.
We observe this well-known behavior in Fig.~\ref{fig:colloidprofiless1}; close to the critical 
point the profiles are smoother and modulations less pronounced.
Away from the critical point, the interface is sharp but the periodic modulations due to the surface extend to far
in the bulk fluid.
The inset of Fig.~\ref{fig:colloidprofiless1} shows the corresponding polymer profiles.
In Ref.~\cite{trokhymchuk98}, the main result is that the interface widens due to the porous medium.
The same happens here and is due to fact that one is effectively closer to the critical point.

In Fig.~\ref{fig:colloidprofiless50}, as we saw for the bulk phase diagram, 
there is a simple rescaling at work and the profiles merely
differ with a factor $(1-\eta_m)$.
The inset in Fig.~\ref{fig:colloidprofiless50} shows the same colloid profiles but now rescaled 
and we have zoomed in on the region close to the interface.
Clearly, even the modulations follow the case without matrix with the same accuracy
as the bulk coexistence values in the inset of Fig.~\ref{fig:sgcphasediagrams50}.

We have also studied the asymptotic decay of correlations with the interface via the density profiles.
These must be of the same nature as the decay of the direct correlations in bulk (determined via the ROZ equations, 
see previous subsection), i.e.\ either monotonic or periodic~\cite{evans94}.
However, determining the crossing points of the FW line with the binodals using the interfacial
profiles yields a systematic shift away from the critical point, compared to the bulk calculation ($\sim 5\%$).
Probably, this is due to numerical limits.
Close to this crossing point both (the periodic and the monotonic) modes of decay are equally strong, 
so only far away from the interface truly asymptotic behavior may be observed.
However, there, the periodic modulations may have become too small to be observable.
Furthermore, our numerical routine has no real incentive to minimize the tails of the profiles
as the gain in free energy is very low.
\begin{figure}[t]
\centering
\epsfig{figure=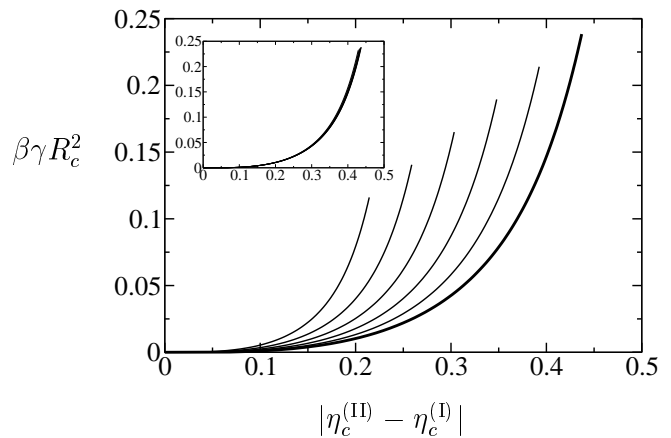,width=8.7cm}
\caption{\small Same as in Fig.~\ref{fig:surfacetensions1} but now for $s=50$.
Again, matrix packing fractions increase from right ($\eta_m=0$, thick curve) to left,
$\eta_m= 0, 0.1, 0.2, 0.3, 0.4, 0.5$.
Each curve is computed from the critical point, $\eta_{p,r}=\eta_{p,r}^{({\rm crit})}$
(where $\gamma =0$) until twice the critical fugacity,
$\eta_{p,r}=2\eta_{p,r}^{({\rm crit})}$.
Inset: the same (but rescaled) curves are shown, i.e.\ $\beta\gamma R_c^2/(1-\eta_m)$ vs.\
$|\eta_c^{({\rm II})}-\eta_c^{({\rm I})}|/(1-\eta_m)$.
}
\label{fig:surfacetensions50}
\end{figure}

\subsection{Fluid-Fluid Surface Tension inside a Bulk Porous Matrix}

The presence of the matrix also affects the surface tension between the colloidal liquid
and gas phases.
The interfacial or surface tension $\gamma$ of planar interfaces in the grand canonical ensemble 
is defined through
\be
\gamma A = \Omega_{\rm inh} + PV
\ee
where $A$ is the amount of surface area, $\Omega_{\rm inh}$
is the grand potential for the inhomogeneous system and $P$ the pressure (i.e.\ $-PV$ is the
grand potential for the homogeneous bulk system).
With our numerical scheme we calculate density profiles in $z$-direction
so it makes sense to write the surface tension as an integral,
\be
\gamma=\int dz\left[ \omega (z)+P\right]
\ee
with
\begin{multline}
\omega (z)=k_{\rm B}T\sum_{i=c,p}\rho_i(z)\left[\ln\left(\rho_i(z)\Delta_i\right)-1\right] \\ 
-\sum_{i=c,p}\mu_i\rho_i(z)+ k_{\rm B}T \Phi (\{ n^i_{\nu}(z)\}).
\end{multline}
The quantity $\omega(z)$ is a ``local'' grand potential density whose average
over space yields the actual grand potential per unit of volume $\Omega_{\rm inh}/V$.
In Figs.~\ref{fig:surfacetensions1} and~\ref{fig:surfacetensions50} we have plotted the
surface tension versus the colloidal density difference in the two phases 
for $s=1$ and $s=50$, respectively.
In both cases the effect of the matrix is that the surface
tensions increase faster with the difference $|\eta_c^{({\rm II})}-\eta_c^{({\rm I})}|$
which is of course due to the fact that the coexistence area becomes less wide
as the coexisting packing fractions themselves become smaller.
In the inset of Fig.~\ref{fig:surfacetensions50}, we show the same curves
rescaled with $(1-\eta_m)$, and the rescaled graphs fall almost on top of the original one without any matrix.
Here, we note that also the surface tension has been rescaled with $(1-\eta_m)$; this
is needed from Eq.~(\ref{eq:rescaling}) as the free energy density $\omega(z)$ needs to be rescaled as well.
Again, this rescaling procedure is not successful for $s=1$.

Often, the surface tension is plotted against the relative distance to the critical point,
$(\eta_{p,r}/\eta_{p,r}^{({\rm crit})}-1)$~\cite{brader02swet}.
However, this does not improve the rescaling for $s=1$ and this can be seen from the fact that
the end points of the curves in Figs.~\ref{fig:surfacetensions1} and~\ref{fig:surfacetensions50}
are all at twice the critical fugacity, $\eta_{p,r}=2\eta_{p,r}^{({\rm crit})}$ and 
the surface tensions (rescaled or not) are at quite different values at the end points.

\section{Conclusion}

We have considered the full ternary system of hard spheres and ideal polymers
(represented by the AO model) in contact with a quenched hard sphere fluid acting as a porous matrix.
Using a QA DFT in the spirit of Rosenfeld's
fundamental measure approach, we studied capillary condensation in a tiny sample of matrix as well as the
the fluid-fluid interface inside a bulk matrix.
The results have been presented in terms of two types of matrices: (i) colloid-sized
matrix particles (size ratio $s=1$) being a reference system
and (ii) matrix particles which are much larger than the colloids (size ratio $s=50$).
The case of small matrix particles is limited to relatively low packing fractions ($\eta_m\sim 0.2$),
whereas in the second case, much higher matrix packing fractions are accessible ($\eta_m\sim 0.5$), the
pores of the matrix being much larger.
Additionally, we have suggested that case (i) as well as (ii) 
could in principle be realized experimentally in 3D, i.e.\ using laser tweezers and
colloidal sediments respectively, to serve as a model porous medium for colloidal suspensions.

We have shown that in the limit of infinitely large matrix particles, the standard AO results 
(without matrix) are recovered via a simple rescaling.
In case of $s=50$ our bulk but also the interface results can be mapped onto the case without matrix
with high accuracy.
However, in the case of small matrix particles ($s=1$) this mapping fails,
which is due to the more complex (and smaller) pore geometry on the colloidal scale.

Assuming a more ``experimental'' point of view, we have considered a tiny sample of porous matrix
immersed in a large system of colloid-polymer mixture.
When the fluid-fluid binodal is approached in the colloid-poor region of the phase diagram,
capillary condensation occurs in the sample.
This transition appears as a capillary line in the phase diagram (in system representation) 
extending along the binodal and ending in a capillary critical point.
In case of small matrix particles, the capillary lines (for various densities of the matrix) 
are well separated from the bulk binodal but the capillary critical points lie deep into the 
colloidal gas regime.
Concerning the large matrix particles, these capillary critical points are located closer to the bulk
critical point, however, the capillary lines are also very close to the binodal.
Still, using density-matched colloidal suspensions, we argue that capillary condensation
may be observable in experiments.

We have computed fluid-fluid profiles inside the porous matrix as well as the corresponding surface tensions.
For $s=50$, these can be mapped onto the case without matrix but for $s=1$
the critical point shifts to higher polymer fugacities.
Therefore, increasing the density of the matrix, profiles become smoother due to effective approach of
the critical point.
Solving the ROZ equations, we have also determined the crossover between monotonic and periodic decay of 
pair correlations of the mixture inside the matrix for $s=1$.
Comparing these with the decay of the interfacial correlations we find a small discrepancy
which is probably due to numerical limits.

It should be noted that we do not expect our current approach to satisfactory describe the
(subtle) phenomena associated with wetting of the curved surfaces of the matrix particles by 
the colloidal liquid~\cite{evans03,evans03a}.
Especially for $s=50$ close to the critical point in the complete wetting regime (of the planar hard wall),
we can well imagine that the growths of thick wetting films preempt our capillary condensation transitions
as well as disturb the fluid-fluid interfaces.

Concerning the fluid profiles, we have only considered a homogeneous background of matrix particles in this paper.
It would be interesting to use inhomogeneous matrix realizations, as e.g.\ a step function
of zero and nonzero matrix packing fraction (i.e.\ the interface of empty space-matrix) or a constant matrix 
background in contact with a hard wall.
Both types could give rise to interesting and substantially modified wetting behaviour.
Additionally, one could also consider other types of matrices, e.g.\ quenched polymers
or combinations of quenched colloids with quenched polymers~\cite{schmidt02pordf,schmidt02aom}.
These are maybe less realistic from an experimental point of view but still interesting
due to the competition of capillary condensation with evaporation.

As we have mentioned in the introduction, there are no experiments concerning phase behavior of 
colloidal suspensions in contact with 3D porous media to our knowledge.
We hope that the accumulating 
results~\cite{cruzdeleon98,cruzdeleon99,kluijtmans97,kluijtmans99,kluijtmans00,weronski03,schmidt02pordf,schmidt02aom}, 
including those in this paper, may encourage more experimental efforts in that direction.
It is important to keep in mind that a suitable porous matrix is a compromise
between length scales: large enough to allow penetration of the colloids into the void space, but small enough
to retain significant surface and capillary effects.
In colloidal fluids in general, these last-mentioned effects are known to be 
much smaller than in atomic systems, thus  providing a formidable challenge to experimentalists
aiming to observe e.g.\ capillary condensation of a colloidal suspension in a porous matrix.

\begin{acknowledgments}
The authors would like to thank D.\ G.\ A.\ L.\ Aarts for useful discussions and
R.\ Blaak for a critical reading of the manuscript.
MS thanks D.\ L.\ J.\ Vossen for detailed explanantions of the laser-tweezer setup and
T.\ Gisler for pointing out Refs.~\cite{cruzdeleon98,cruzdeleon99}.
This work is financially supported by the SFB-TR6 program ``Physics of colloidal dispersions in external fields'' 
of the \emph{Deutsche Forschungsgemeinschaft} (DFG).
The work of MS is part of the research program of the \emph{Stichting voor Fundamenteel Onderzoek der Materie}
(FOM), that is financially supported by the \emph{Nederlandse Organisatie voor Wetenschappelijk Onderzoek}
(NWO).
\end{acknowledgments}

\appendix

\section{Bulk Fluid Free Energy}

The bulk free energy of the colloid-polymer mixture in contact with a homogeneous hard-sphere porous matrix 
as given in Eq.~(\ref{eq:bulkfreeenergydensity}) is
\begin{multline}
f(\eta_c,\eta_p,\eta_m)=\eta_c\left(\ln\eta_c-1\right)+\frac{\eta_p}{q^3}\left(\ln\eta_p-1\right)
\\ +f_0(\eta_c,\eta_m)-\frac{\eta_p}{q^3}\ln\alpha(\eta_c,\eta_m)
\label{eq:bulkfreeenergydensityapp}
\end{multline}
We note the occurence of only third and lower powers of $1/s$
in both $f_0$ and $\alpha$ which are given by
\begin{widetext}
\begin{multline}
f_0(\eta_c,\eta_m)=
\frac{\eta_m}{s^3}\ln(1-\eta_m)-\left(\frac{\eta_m}{s^3}+\eta_c\right)\ln(1-\eta_c-\eta_m) 
+\frac{3\eta_c\eta_m^2 (2+\eta_c(\eta_m-2)-2\eta_m)}{2s^3(1-\eta_m)^2(1-\eta_c-\eta_m)^2} 
\\ +\frac{3\eta_c}{2s^3(1-\eta_c-\eta_m)^2}\left\{ 
\eta_m(2-2\eta_c+\eta_m)s 
+\eta_m(2+\eta_c-2\eta_m)s^2
+\eta_c(2-\eta_c-2\eta_m)s^3
\right\}
\end{multline}
and
\begin{multline}
\ln\alpha(\eta_c,\eta_m)=\ln(1-\eta_c-\eta_m)+ \\
-\frac{q}{2s^3(1-\eta_c-\eta_m)^3}
\left\{
2\eta_m(1+\eta_c^2+\eta_m+\eta_m^2-\eta_c(2+\eta_m))q^2 \right. \\
\left.
-3\eta_m q \left[ -2+\eta_m+\eta_m^2+2\eta_c^2(-1+q)-\eta_c(-4+\eta_m+2q+4\eta_m q) \right] s \right. \\
\left. +
6\eta_m\left[(1-\eta_c-\eta_m)^2+3\eta_c(1-\eta_c-\eta_m)q+\eta_c(1+2\eta_c-\eta_m)q^2\right]s^2\right. \\
\left. +
\eta_c\left[-\eta_c(1-\eta_m)(12+(3-2q)q)+2(1-\eta_m)^2(3+q(3+q))+\eta_c^2(6+q(-3+2q)) \right]s^3
\right\} .
\end{multline}
\end{widetext}


\end{document}